\documentclass[a4paper,12pt]{article}
\newcommand{\sect}[1]{\setcounter{equation}{0}\section{#1}}

\newcommand{\EQ}{\begin{equation}}
\newcommand{\EN}{\end{equation}}
\newcommand{\bea}{\begin{eqnarray}}
\newcommand{\ena}{\end{eqnarray}}
\newcommand{\vs}[1]{\vspace{#1 mm}}

\renewcommand{\a}{\alpha}

\renewcommand{\d}{\delta}
\newcommand{\e}{\epsilon}
\def\bbox{{\,\lower0.9pt\vbox{\hrule \hbox{\vrule height 0.2 cm
\hskip 0.2 cm \vrule height 0.2 cm}\hrule}\,}}
\newcommand{\dsl}{\pa \kern-0.5em /}

\newcommand{\pa}{\partial}
\renewcommand{\t}{\theta}

\newcommand{\nn}{\nonumber\\}
\newcommand{\p}[1]{(\ref{#1})}
\newcommand{\lan}{\langle}
\newcommand{\ran}{\rangle}
\begin{document}

\topmargin 0pt
\oddsidemargin 0mm

\renewcommand{\thefootnote}{\fnsymbol{footnote}}
\begin{titlepage}

\setcounter{page}{0}
\begin{flushright}
OU-HET 358 \\
hep-th/0009021
\end{flushright}

\vs{10}
\begin{center}
{\Large\bf Noncommutative Gauge Dynamics from Brane Configurations
 with Background $B$ field}
\vs{15}

{\large
Nobuyoshi Ohta\footnote{e-mail address: ohta@phys.sci.osaka-u.ac.jp}
and
Dan Tomino\footnote{e-mail address: dan@het.phys.sci.osaka-u.ac.jp}} \\
\vs{10}
{\em Department of Physics, Osaka University,
Toyonaka, Osaka 560-0043, Japan}

\end{center}
\vs{15}
\centerline{{\bf{Abstract}}}
\vs{5}

We study $D=3$ field theories on the D3-branes stretched between two
NS5-branes with NS $B$-field background. The theory is a noncommutative
gauge theory. The mirror symmetry and S-duality of the theory are discussed.
A new feature is that the mirror of the noncommutative gauge theory is not
a field theory but an open string decoupled from the closed string.
We also consider brane creation phenomena and use the result to discuss
the analogue of the Seiberg duality.
A noncommutative soliton is interpreted as a D1-brane induced on D3-brane.

\end{titlepage}
\newpage
\renewcommand{\thefootnote}{\arabic{footnote}}
\setcounter{footnote}{0}

\sect{Introduction}

The last few years have witnessed a remarkable progress in our understanding
of nonperturbative properties of supersymmetric Yang-Mills theories (SYM) in
various dimensions. Many interesting and exact results have been obtained
in refs.~\cite{IS,BHO1}. These results were interpreted in terms of the
field theory realized on the worldvolume of the brane configurations
in type IIB superstring theory~\cite{PZ}-\cite{GK}. In particular,
the miraculous ``mirror symmetry'' from the field theory point of view
can be simply understood as an $SL(2,{\bf Z})$ duality symmetry in type
IIB superstring theory.

Four-dimensional theories can be also discussed by using brane configurations
in type IIA superstring~\cite{EGK}-\cite{W1}. The so-called Seiberg
duality~\cite{S} can be easily understood~\cite{EGK} in terms of brane
exchange and creation phenomena first discussed in ref.~\cite{HW}. These
theories have an interpretation in terms of M-brane configuration~\cite{W1}.
For a review of these developments, see ref.~\cite{GK}. In fact, other BPS
brane configurations give various field theories including the above-mentioned
ones upon compactifications and T-dualities, and it is possible to classify
brane configurations~\cite{OT,KOO}.

Recently it has been noticed that noncommutative SYM (NCSYM) has also
a similar realization on the worldvolume of the branes if we include
background NS $B$-field~\cite{DH}-\cite{SW}. This naturally leads us to
consider various brane configurations discussed so far in the presence of the
background $B$ field. NCSYM in type IIA superstring has been considered in
ref.~\cite{SJ}. The purpose of this paper is to discuss $D=3$ theories on
brane configuration of Hanany-Witten~\cite{HW} and its variant.

In what follows we first study the $D=3$ NCSYM on D3-branes stretched between
two parallel NS5-branes with NS $B$-field background in type IIB superstring.
We show the type IIB $SL(2,{\bf Z})$ duality symmetry tells us that the
mirror of the NCSYM is not just a field theory but noncommutative open string
theory (NCOS)~\cite{SST,GMMS}. We also consider brane creation phenomena
and use the result to discuss the $D=3$ NCSYM version of the Seiberg
duality. Our analysis suggests that the analogue of the Seiberg duality is
valid also in $D=3$ NCSYM.

The paper is organized as follows. In sect.~2, we briefly review how
noncommutative theory is realized on the worldvolume of the branes in the
presence of background NS $B$-field. The BPS properties of various brane
configurations are discussed in sect.~3. Properties of $D=3$ SYM such as
mirror symmetry are discussed in sect.~4. In sect.~5, we show that the brane
configuration suggests the validity of the NCSYM version of the Seiberg
duality. Sect.~6 is devoted to discussions.

\sect{Field theory on D-branes with background NS $B$-field}

In this section we briefly review how noncommutativity appears in the theories
realized on D-branes with NS $B$-field background.

The bosonic part of the open string action is given by
\bea
S=- \frac{1}{4\pi\a'} \int_{\Sigma} ( g_{ij} \pa_{a}X^{i} \pa^{a}X^{j}
 - 2\pi\a' B_{ij} \e^{ab} \pa_a X^i \pa_b X^j),
\label{act}
\ena
where $\Sigma$ is the string worldsheet. The variation of the action~\p{act}
leads to the boundary conditions
\bea
g_{ij}\pa_{\sigma}X^{j} + 2\pi\a' B_{ij}\pa_{t}X^{j}\Big|_{\pa\Sigma} = 0,
&& {\rm for} \;\; i,j:{\rm worldvolume \; directions}, \nn
g_{mn}\pa_{t}X^{n} \Big |_{\pa\Sigma} = 0, && {\rm for}\;\;
 m,n :{\rm otherwise}.
\label{bc}
\ena
Note that the first line reduces to the usual Neumann condition in the
absence of the $B$ field. The additional term induces the mixing between
the ``momentum'' and coordinates and gives rise to noncommutativity.
The canonical momenta are given by
\bea
P_i = \frac{1}{2\pi \a'} (g_{ij} \pa_t X^j + 2\pi \a' B_{ij} \pa_\sigma X^j).
\label{mom}
\ena

Imposing the boundary condition~\p{bc}, the mode expansions for fixed $i$
and $j$ are given as
\bea
X^i &=& x^i + 2\a' (g^{ij}p_j t - 2\pi \a' B^{ij} p_j \sigma) \nn
&& + \sqrt{2\a'} \sum_{n\neq 0} \frac{e^{-int}}{n} ( ig^{ij}\a_{jn}
 \cos n\sigma -2\pi \a' B^{ij} \a_{jn} \sin n\sigma), \nn
X^j &=& x^j + 2\a' (g^{ij}p_i t + 2\pi\a'B^{ij} p_i \sigma) \nn
&& + \sqrt{2\a'} \sum_{n\neq 0} \frac{e^{-int}}{n} ( ig^{ij}\a_{in}
 \cos n\sigma +2\pi\a' B^{ij} \a_{in} \sin n\sigma).
\label{me}
\ena
An interesting property of the mode expansions is that there is a
noncommutativity in the coordinates in \p{me} as indicated in the mixing of
the momenta and coordinates, but the momenta given in \p{mom} have the
expansions
\bea
2\pi \a' P_i &=& \left[g^2+ (2\pi \a' B)^2\right]_{ij} \left[2\a' p^j
 + \sqrt{2\a'} \sum_{n} e^{-int} i\a^j_n \cos n\sigma \right], \nn
2\pi \a' P_j &=& \left[g^2+ (2\pi\a' B)^2\right]_{ij} \left[2\a' p^i
 + \sqrt{2\a'} \sum_{n} e^{-int} i\a^i_n \cos n\sigma \right],
\label{mom2}
\ena
and commute with each other.

Imposing the equal-time commutation relations
\bea
[ X^i(\sigma, t), P_j(\sigma', t) ] = i {g^i}_j \d(\sigma-\sigma'),
\ena
and that $[X^i(\sigma, t), X^j(\sigma', t)]$ (almost) vanish, we get
\bea
[ x^i, x^j ] = i \t^{ij}, \quad
[ x^i, p^j ] = i G^{ij}, \quad
{}[ \a^i_n, \a^j_m ] = n \d_{n+m,0} G^{ij},
\label{cr}
\ena
where $G^{ij}$ and $\t^{ij}$ are the open string metric and
noncommutativity parameter defined by
\bea
G^{ij} &=& \left( \frac{1}{g + 2\pi\a'B} \right)^{ij}_{S}, \nn
G_{ij} &=& g_{ij} - (2\pi\a')^{2}(Bg^{-1}B)_{ij}, \nn
\t^{ij} &=& 2\pi\a' \left(  \frac{1}{g + 2\pi\a'B}\right)^{ij}_{A},
\label{osm}
\ena
with $(\,)_{S}$ and $(\,)_{A}$ denoting the symmetric and antisymmetric parts.
One then finds
\bea
[ X^i(\sigma, t), X^j(\sigma', t)] = \left\{
\begin{array}{ccl}
i \t^{ij} & {\rm for} & \sigma =\sigma'= 0, \\
-i \t^{ij} & {\rm for} & \sigma=\sigma'= \pi, \\
0 & {\rm otherwise}. &
\end{array} \right.
\ena
Thus the momenta are commutative whereas the coordinates are not at the
boundary.

In terms of the Euclideanized coordinate
\bea
z= e^{\tau + i \sigma}, \quad
\tau \equiv it,
\ena
the mode expansion~\p{me} and commutation relations~\p{cr} give the correlator
\bea
\lan X^{i}(z)X^{j}(z')\ran &=&
- \a' \Big[ g^{ij}\log|z-z'| - g^{ij} \log|z - \bar{z}'| \nn
&& + G^{ij}\log|z - \bar{z}'|^{2} + \frac{1}{2\pi\a'} \t^{ij}
\log \frac{z - \bar{z}'}{\bar{z} - z'} + D^{ij} \Big ], \nn
\langle X^{m}(z)X^{n}(z')\rangle &=&
- \a' \Big[ g^{mn}\log|z-z'| - g^{mn}\log|z - \bar{z}'|\Big ].
\label{cor}
\ena
where $D^{ij}$ is a constant that may depend on $B$ but not on $z$ and
$\bar{z}$. On the boundary at $\sigma=0,\pi$, the correlator reduces to
\bea
\lan X^{i}(\tau)X^{j}(\tau')\ran =
- \a' G^{ij}\log(\tau - \tau')^{2} + \frac{i}{2} \t^{ij} \e(\tau - \tau'),
\label{corb}
\ena
which gives rise to the star product in the interactions between open
string states given by vertex operators~\cite{SW}.

In the scaling limit
\bea
\a' &\sim& \e^{1/2} \to 0, \nn
g_{ij} &\sim& \e \to 0,
\label{sl}
\ena
the massive and closed string modes decouple and the effective theory of
the remaining massless modes on the D$p$-brane is NCSYM theory~\cite{SW}.
Though we have discussed the effective theory only for the bosonic part,
once we identify the unbroken supersymmetry in the brane configuration,
it is easy to construct the corresponding NCSYM.

Our next task is to identify the BPS brane configurations and which massless
modes are kept in the resulting theory.

\sect{BPS brane configurations}

In this section, we analyze BPS brane configurations with $B_{12} \neq 0$.
The configurations we have in mind are basically the same as those in
ref.~\cite{HW}.

\subsection{NS5-D5-D3 system}

We first consider type IIB theory in ten-dimensional flat Minkowski space
with trivial background fields. The brane configuration of our interest is
composed of D3-branes suspended between two parallel NS5-branes with D5-branes
for additional matter hypermultiplets. In the flat space, an NS5-brane with
the worldvolume in the (012345) directions is invariant under half of the
supersymmetries determined by
\bea
\e_{L} = \Gamma_{012345} \e_{L}, \qquad
\e_{R} = -\Gamma_{012345} \e_{R}.
\label{ns5}
\ena
Note that the spinors $\e_{L}$ and $\e_{R}$ satisfy $\Gamma \e_{L} = \e_{L},
\Gamma \e_{R} = \e_{R}$ for $\Gamma = \Gamma_{012\cdots9}$. Here we use the
notation $\Gamma_{012\cdots p}$ for antisymmetrized product of gamma matrices
with unit strength.

On the other hand, a D5-brane with the worldvolume in the (012789) directions
is invariant under half of the supersymmetries given by the different
condition
\bea
\e_{L} = \Gamma_{012789}\,\, \e_{R}.
\label{d5}
\ena
Each of these conditions \p{ns5} and \p{d5} preserves 16 components of
supersymmetry generators. If there are both the NS5- and D5-branes in flat
space, only the supersymmetry compatible with the above two conditions
remains unbroken. We thus find that only eight components of the supersymmetry
generators are preserved, giving a 1/4 BPS state.

It is possible to add another D3-brane with the worldvolume (0126) to this
system without breaking supersymmetry completely. The additional condition
we get is
\bea
\e_{L} = \Gamma_{0126} \e_{R},
\label{d3}
\ena
which is automatic given the conditions \p{ns5} and \p{d5}~\cite{HW}.

Next let us introduce a constant NS $B$-field background, $B_{12} = B$
and discuss BPS brane configurations. To examine how the condition~\p{d3}
will be modified by the presence of $B$ field, we go back to the boundary
condition~\p{bc} for D$p$-brane in the presence of $B$ field. For our
D3-brane, this becomes
\bea
\pa_{\sigma}X^{0,6} &=& 0, \nn
\pa_{\sigma}X^{1} + 2\pi i\a' B \pa_{t}X^{2} &=& 0, \nn
\pa_{\sigma}X^{2} - 2\pi i\a' B \pa_{t}X^{1} &=& 0, \nn
\pa_{t}X^{3,4,5,7,8,9} &=& 0.
\label{d3bc}
\ena
It is then useful to consider the T-dual transformation in the $X^{2}$
direction~\cite{SJ} and note that the T-dual transformation exchanges
$\pa_{\sigma}$ with $-i\pa_{t}$. We are thus lead to
\bea
\pa_{\sigma}X^{0,6} &=& 0, \nn
\pa_{\sigma}\left( \frac{1}{\sqrt{1+(2\pi\a' B)^2}}X^{1}
- \frac{2\pi\a' B}{\sqrt{1+(2\pi\a' B)^2}}X^{2}\right) &=& 0, \nn
\pa_{t} \left(\frac{1}{\sqrt{1+(2\pi\a' B)^2}}X^{2} + \frac{2\pi\a' B}
{\sqrt{1+(2\pi\a' B)^2}} X^{1}\right) &=& 0, \nn
\pa_{t}X^{3,4,5,7,8,9} &=& 0.
\label{d3bc1}
\ena
This is a D2-brane boundary condition rotated in the (12)-plane by the
angle given by $\tan \t_{12} = -2\pi \a'B$. For such rotation,
the gamma matrix $\Gamma_{1}$ transforms to $e^{\frac{1}{2}\t_{12}
\Gamma_{12}}\Gamma_{1}e^{-\frac{1}{2}\t_{12}\Gamma_{12}} = e^{\t_{12}
\Gamma_{12}}\Gamma_{1}$. Thus the unbroken supersymmetry of this D2-brane
is given by
\bea
\e_{L} = e^{\t_{12}\Gamma_{12}}\Gamma_{016}\e_{R}.
\label{ubs}
\ena
Making T-duality transformation back in the $X^{2}$ direction, we obtain
unbroken supersymmetry of D3-brane in the constant background determined by
\bea
\e_{L} = e^{\t_{12}\Gamma_{12}}\Gamma_{0126}\e_{R}.
\label{d3bc2}
\ena

On the other hand, unbroken symmetry of NS5-brane is unaffected.
The condition \p{ns5} comes from the requirement that the supersymmetry
transformation on the NS5-brane action be zero~\cite{BKO}, but
this vanishes for constant NS $B$ field and \p{ns5} does not change~\cite{SJ}.

We now have to examine the remaining supersymmetry under \p{ns5} and
\p{d3bc2}. The easiest way to identify this is to redefine spinors by
\bea
\e_R'= e^{\frac{1}{2}\t_{12}\Gamma_{12}}\e_R, \qquad
\e_L'= e^{-\frac{1}{2}\t_{12}\Gamma_{12}}\e_L.
\ena
We then find that the conditions \p{ns5} and \p{d3bc2} reduces back to
\p{ns5} and \p{d3} without $B$ in terms of $\e_{L,R}'$. We thus conclude
that eight supersymmetries are still unbroken for NS5- and D3-branes
even in the presence of constant $B$. Combining \p{ns5} and \p{d3bc2},
we can show that the D5-branes are also allowed in the configuration without
breaking any supersymmetry.

In summary, we have shown that the configuration with NS5-, D5- and D3-branes
keeps eight supersymmetries unbroken in constant background fields
(as in the absence of $B$ field~\cite{HW}).

\subsection{D0 - D4 system \, --- \, instanton}

As a simple application of the above discussions and check of our method,
let us consider a BPS D0-D4 system.

A type IIA configuration with D0- and D4(01234)-branes is a 1/4 BPS
configuration in flat Minkowski space without $B_{ij}$. Now, we consider
D0- and D4-branes in a constant background: $g_{ij} = \eta_{ij}, B_{12}\neq 0,
B_{34} \neq 0, B_{ij \neq 1234} = 0$. Unbroken supersymmetry is given by
\bea
\e_{L} &=& e^{\t_{1}\Gamma_{12} + \t_{2}\Gamma_{34}}\Gamma_{01234}\e_{R}, \nn
\e_{L} &=& \Gamma_{0}\e_{R},
\ena
where the angles are given by $\tan\t_{1} = -2\pi\a' B_{12}, \tan\t_{2}
= -2\pi\a' B_{34}$. The first (second) line comes from D4(D0)-brane.
The compatibility of these condition requires
\bea
\e_{L} &=&e^{\t_{1}\Gamma_{12} + \t_{2}\Gamma_{34}}\Gamma_{01234}
(\Gamma_{0})^{-1} \e_{L} \nn
&=& e^{\t_{1}\Gamma_{12} + \t_{2}\Gamma_{34}}\Gamma_{1234} \e_{L}.
\label{com}
\ena
Thus the number of eigenvalues 1 of the matrix $e^{\t_{1}\Gamma_{12} +
\t_{2}\Gamma_{34}}\Gamma_{1234}$ gives the number of unbroken supersymmetries.
Now $\Gamma, \Gamma_{1234}, i\Gamma_{12}$ and $i\Gamma_{34}$ commute with
each other and square to the identity and the traces of their products vanish.
So we can diagonalize these matrices as
\bea
\Gamma &=& {\bf 1}_{16} \otimes \sigma_{3}, \nn
\Gamma_{1234} &=& {\bf 1}_{8} \otimes \sigma_{3} \otimes {\bf 1}_{2}, \nn
\Gamma_{12} &=& i{\bf 1}_{4}\otimes \sigma_{3}\otimes {\bf 1}_{4}, \nn
\Gamma_{34} &=& -\Gamma_{12}\,\Gamma_{1234} \nn
&=& -i{\bf 1}_{4}\otimes \sigma_{3} \otimes \sigma_{3} \otimes {\bf 1}_{2},
\label{mat}
\ena
where ${\bf 1}_n$ is an $n\times n$ unit matrix. We then find
\bea
e^{\t_{1}\Gamma_{12} + \t_{2} \Gamma_{34}}\Gamma_{1234}
&=& {\bf 1}_{4} \otimes \pmatrix{ e^{i(\t_{1}-\t_{2})}& & & \cr
  & e^{-i(\t_{1}-\t_{2})}& & \cr
  & & -e^{i(\t_{1}+\t_{2})}& \cr
  & & & -e^{-i(\t_{1}+\t_{2})} \cr} \otimes {\bf 1}_{2},\nn
\label{cond1}
\ena
We must take the fact into account that the spinor $\e_{L}$ satisfies the
additional condition $\Gamma \e_{L} = \e_{L}$, reducing the number of
supersymmetry by half. We then find that in the presence of $B$ field,
supersymmetry is unbroken partially if $\t_{1}= \t_{2}$ {\it i.e.}
$B_{12} = B_{34}$ or $\t_{1}+\t_{2}=\pi$ {\it i.e.} $B_{12} = -B_{34}$, and
in both cases eight supersymmetries are preserved. Our result is consistent
with that in ref.~\cite{SW}.

\sect{NCSYM on the branes, brane creation and mirror symmetry}

In this section, we identify the field theory on the worldvolume of the
D3-branes in the brane configuration and discuss some nonperturbative
properties of NCYM, brane creation and mirror symmetry.

\subsection{Identification of the theory}

It is well known that the effective low-energy theory on the D-branes
are SYM theories. From the discussions in sect.~2, it is clear that in the
presence of the background NS $B$-field, the theory becomes NCSYM.
The supersymmetry on the D3-branes would be $D=3, N=8$, which is reduced
by the two parallel NS5-branes to $D=3, N=4$.

Let us first summarize how the massless degrees of freedom on the D3-branes
are reduced by the boundary conditions imposed by other branes~\cite{HW}.
\begin{enumerate}
\item
For D3-branes stretched between two parallel NS5-branes, the massless
fields on the D3-branes are the vector multiplets $A_\mu, X^3, X^4, X^5$,
because Dirichlet conditions are imposed on the directions $X^6, \ldots,
X^9$.
\item
For D3-branes stretched between two D5-branes, the massless fields are
the hypermultiplets $X^6, X^7, X^8, X^9$, because Dirichlet conditions
are imposed on the directions $X^3, \ldots,X^5$.
\item
For D3-branes stretched between an NS5-brane and a D5-brane, there remains
no massless modes.
\end{enumerate}

As a result, the effective theory on the D3-branes stretched between two
parallel NS5-branes in the constant NS $B_{12}$ background is a $D=3, N=4$
NCSYM theory with $N=4$ vector multiplets $A_\mu, X^3,X^4,X^5$. If we add
D5-branes to this system, we get $N=4$ matter hypermultiplets
$X^6,X^7,X^8,X^9$. This is all the same as~\cite{HW}, except that the theory
is noncommutative in $x^1$ and $x^2$.

We can interpret geometrical quantities in these configurations as
informations on the moduli space of NCSYM on D-branes, just as in
ref.~\cite{HW}, at least at the classical level. For example, positions
of D3-branes are interpreted as the vevs of the scalar fields which
parametrize Coulomb and Higgs branches of the $D=3, N=4$ NCSYM, and the
length of D3-branes in the $X^{6}$ direction, which is limited by the two
NS5-branes, is proportional to the inverse of the gauge coupling
constant $1/g^{2}_{YM}$ of NCSYM.

There also exist some new aspects such as noncommutative solitons~\cite{GMS},
noncommutative monopole~\cite{HI}, and noncommutative instanton~\cite{NS}.
It would be interesting to give a brane realization of these solitonic
solutions as in ref.~\cite{LLO}. This will be discussed briefly in sect.~6.

\subsection{Brane creation}

In these brane configurations, matter fields in the fundamental representation
in NCSYM come from the open strings stretching between D3-branes and the
D5-branes perpendicular to the two NS5-branes. The length of the
open strings gives the bare mass to matter fields and this is interpreted as
a moduli of NCSYM. The moduli are $X^{3},X^{4},X^{5}$ for D5-branes. It is
important here to realize that $X^{6}$ is irrelevant parameter for NCSYM
and the theory is not affected by this position. This observation can be
used to argue that brane creation must occur~\cite{HW} in this system.

The reason is the following. First let us consider $n$ D3-branes between
the two NS5-branes and denote the positions of the two NS5-branes in the
$X^6$ direction by $t^{1}$ and $t^{2}$. The D3-brane worldvolume is limited
in the $X^6$ direction to $t^{1}\le X^6 \le t^{2}$. Now suppose that there
are $m$ D5-branes located in the range $t^{1}\le X^6 \le t^{2}$. The
low-energy effective theory on the D3-branes in this system is $U(n)$
NCSYM with $m$ hypermultiplets.

Next if we move D5-branes in the $X^6$ direction away to the region
$X^6 < t^{1}$ or $t^{2}< X^6$, the lengths of the open strings stretching
between D3- and D5-branes cannot be zero and hence no massless hypermultiplet
does not seem to be produced. However, this cannot be the case since the
effective theory does not depend on the position of the D5-branes. This leads
us to conclude that a new D3-brane must be created between NS5- and D5-branes
when the D5-brane crosses through the NS5-brane in the $X^6$ direction. The
hypermultiplets then arise from the open strings stretching between the new
D3-branes and $n$ D3-branes. This process is depicted in Fig.~1.
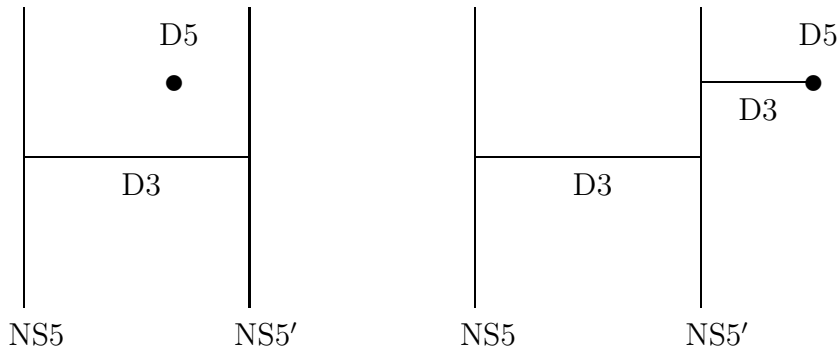
\begin{figure}[htbp]
\setlength{\unitlength}{1mm}
\begin{picture}(50,50)(-30,-5)
\multiput(0,0)(60,0){2}{\line(0,1){40}}
\multiput(30,0)(60,0){2}{\line(0,1){40}}
\multiput(0,20)(60,0){2}{\line(1,0){30}}
\multiput(20,30)(85,0){2}{\circle*{2}}
\put(90,30){\line(1,0){15}}
\multiput(-2,-5)(60,0){2}{NS5}
\multiput(28,-5)(60,0){2}{NS5$'$}
\multiput(18,35)(85,0){2}{D5}
\multiput(13,15)(60,0){2}{D3}
\put(95,25){D3}
\end{picture}
\caption{Brane creation}
\end{figure}

Even in the existence of $B$, the above reasoning does not change.
It follows from the same argument that branes must be created.
The effect of purely spatial $B$ field is essentially to rescale the
momentum and the structure of the string massless modes is not affected
(see \p{mom2}). This is the basic reason why there is no difference in the
brane creation even with the $B$ background. For consistency, it is expected
that the s-rule is also valid.

Though the consistency of the field theory on the brane requires the
brane creation, it would be interesting to confirm this phenomenon by more
explicit calculations like those in refs.~\cite{BDG,DFK,BGL,KOZ}.

\subsection{Mirror symmetry}

It is known that many $D=3, N=4$ SYM theories have dual description in terms
of those called mirror theories at the IR fixed point~\cite{IS}. This
phenomenon was understood in ref.~\cite{HW} by a series of operation in the
brane system. It consists of an S-dual transformation and a rotation that
maps $X^{j}$ to $X^{j+4}$ and $X^{j+4}$ to $-X^{j}$ for $j=3,4,5$, and this
amounts to the exchange of the NS5- and D5-branes. Under this operation,
geometrical properties of the brane system do not get modified. The moduli
spaces of the resulting two effective theories interchanged under this
mirror operation have the same structure. This is what is called mirror
symmetry in ref.~\cite{HW}.

Now we wish to consider the mirror symmetry for NCSYM. However, the mirror
operation involves the S-dual transformation, under which NS $B$ field is
mapped to RR 2-form field $C$. Unfortunately little is known about the
effective theory in RR field backgrounds, and it is difficult to identify
what is the resulting theory.

It has been pointed out in ref.~\cite{GMMS}, however, that the NS $B$-field
background can be transformed to that of the gauge field strength on the
worldvolume and then one can use the Montonen-Olive duality (electromagnetic
duality) on the D3-brane for the S-duality. The electromagnetic duality
transformation interchange magnetic field $F_{ij \neq 0}$ to electric field
$F_{0i}$. So under this S-duality, the effective theory on the D3-branes in
the background $B_{ij \neq 0}$ is mapped into a theory with the background
$B_{0i}$. String theory in the background $B_{0i}$ is studied in
refs.~\cite{SST,GMMS}. This theory has space-time noncommutativity
$[X^{0},X^{i}] \neq 0$, and the decoupling limit of the gravity can be defined
for a critical value of the background field $B_{0i}$, but there is no field
theory limit. The theory becomes open string with space-time noncommutativity,
the so-called noncommutative open string (NCOS). Thus under the mirror
operation on our brane systems, NCSYM is mapped into NCOS. All this result is
also confirmed from the dual supergravity description~\cite{RS,CO2,LRS}.
Geometrically, we expect that there is no change in the moduli space under
the mirror operation.

However, there remains a small subtlety in our interpretation. The resulting
NCOS theory by the mirror map contains noncommutativity in the
$t,x^6$ directions, and the latter corresponds to the direction limited
by the 5-branes. This imposes a restriction only on the zero mode sector
of the open string modes but not on the non-zero modes. With such restriction
we expect that when there is a NCSYM, there exist NCOS which is a mirror to
NCSYM in the sense of ref.~\cite{IS}.

To summarize, new and interesting aspects of this mirror symmetry are as
follows:
\begin{itemize}
\item
The mirror of a field theory (which is not a usual one but NCSYM)
is not a field theory but NCOS, which is still decoupled from the gravity.
\item
Pure spatial noncommutativity is mapped into space-time noncommutativity.
\item
In ref.~\cite{IS}, it was shown that the mirror symmetry exchanges the
Coulomb and Higgs branches of SYM. Coulomb branch receives quantum corrections
whereas the Higgs branch does not because of nonrenormalization theorems.
Moduli space of the Higgs branch is the same as that of the instantons.
There does not seem to be much change in the moduli space in noncommutative
theory in this regard.
\end{itemize}

\section{An analogue of the Seiberg duality}

The brane configuration discussed so far consists of D3-branes stretched
between two parallel NS5-branes with additional D5-branes for matter
hypermultiplets, and the effective theory is $D=3,N=4$ NCSYM.

The problem we would like to discuss here is another nonperturbative
effect in $D=3, N=2$ SYM theory known as Seiberg duality~\cite{S}.
The corresponding configuration is easy to identify; it is simply given
by rotating one of the NS5-branes~\cite{EGK,BA}. The rotated NS5-brane will
be referred to as NS5$'$-brane. The brane configuration is then given by
$$
{\rm NS5}: (012345), \qquad
{\rm D3}: (0126), \qquad
{\rm D5}: (012789), \qquad
{\rm NS5}': (012389).
$$
It can be checked that the remaining supersymmetry is $D=3, N=2$ in the
presence of background $B$ field.

Another important ingredient in the discussion is the brane creation
phenomena, which we have already seen that indeed it must occur in the
previous section.

Combining these two observations, we can show the NCSYM version of the
Seiberg duality; the $D=3,N=2$ NCSYM with the gauge group $U(N_c)$ and
$N_f$ matter hypermultiplets are equivalent to the NCSYM with the gauge
group $U(N_f-N_c)$ and $N_f$ matter hypermultiplets for $N_f>N_c$.
The brane exchange and creation process considered in ref.~\cite{EGK} was
made in the direction orthogonal to the $B$-field and obviously it is
not modified in the presence of $B$-field. After this process, we find that
the resulting effective theory on the D3-brane worldvolume is a $D=3$ NCSYM
with the gauge group $U(N_f-N_c)$ with $N_f$ matter hypermultiplets, just as
for the ordinary case~\cite{EGK}. This is the desired analogue of the Seiberg
duality.
Though we believe that the above transformation of brane configuration gives
a strong evidence of this duality, it is desirable to check this from the
field theoretic point of view.

\sect{Discussions}

We have constructed BPS brane configurations corresponding to NCSYM
and examined some nonperturbative properties of the theories on the
worldvolume. For the theory on D3-branes suspended between NS5-branes
in type IIB superstring, we find that $D=3, N=4$ NCSYM is realized, and
the mirror theory of this theory turns out to be NCOS. It is extremely
interesting that the mirror of a field theory is not simply a field theory
but an open string.

It is also easy to reduce the number of supersymmetry by rotating one of
the NS5-branes, as can be understood from the results in refs.~\cite{OT,KOO}.
We have used one of these configurations with 1/4 supersymmetry
to give a strong evidence of the $D=3$ NCSYM version of the Seiberg duality.
After a series of brane exchange and creation, we find that the NCSYM theory
on the D3-brane worldvolume with the gauge group $U(N_c)$ with $N_f$ matter
hypermultiplets is equivalent to a NCSYM with the gauge group $U(N_f-N_c)$
with $N_f$ matter, just as for the ordinary case~\cite{EGK}. Though we
believe that the transformation of brane configuration is a strong evidence
of this duality, it is desirable to check this from the field theoretic
point of view.

This type of the brane configuration is usually considered for T-dualized
type IIA configuration of D4 between two NS5's~\cite{EGK}, and gives the
usual Seiberg duality. However, in noncommutative case, there is a problem.
It has been pointed out that the $U(1)$ sector is frozen for the configuration
of $N$ D4-branes and the gauge group is $SU(N)$ but not $U(N)$~\cite{W1}.
However it is believed that a noncommutative $SU(N)$ SYM theory may not be
consistent~\cite{A}. This problem is currently under study.

The above discussions look similar to the ordinary SYM. However, from the
field theory viewpoint, the properties of the NCSYM under mirror symmetry and
Seiberg duality are quite nontrivial. Only with the brane technique and our
identification of the suitable BPS brane configuration is it possible to cast
this difficult problem to a tractable one. It is then important to understand
clearly to what extent the argument remains the same and where it is modified.
Our results on the mirror symmetry and Seiberg duality were not known and are
already nontrivial for {\it noncommutative} SYM. In addition, there are
several crucial new features. First, as emphasized above, the mirror theory
of the NCSYM is not just a field theory but an open string with space-time
noncommutativity decoupled from the closed string.

Another interesting difference is the existence of the nontrivial soliton
states in NC theories~\cite{GMS}-\cite{NS}. It would be interesting to try to
identify such solutions in our brane configurations~\cite{LLO}. In this
connection, it has been suggested that the soliton in noncommutative scalar
field theory~\cite{GMS} can be interpreted as a D-brane~\cite{HK}. Similar
solitonic solutions with constant field strength are considered for NCSYM in
ref.~\cite{GMS}, and we suspect that these can be again interpreted as
D-branes in the NCSYM on the brane configurations.

The Chern-Simons coupling in the effective theory on the D$p$-branes
\bea
\mu_{{\rm D}p} \int_{{\rm D}p} C \wedge 2\pi \a' B
= N_{{\rm D}(p-2)} \mu_{{\rm D}(p-2)} \int_{{\rm D}(p-2)} C,
\ena
produces D$(p-2)$-brane charge $N_{{\rm D}(p-2)}$ for nonzero NS $B$-field
background. This means that $N_{{\rm D}(p-2)}$ D$(p-2)$-branes are induced
on the D$p$-brane. In our $D=3$ theory, using $\mu_{{\rm D}p}=
\frac{\pi}{\kappa^2}(4\pi^2 \a')^{3-p}$, one finds
\bea
N_{{\rm D}1} = \frac{1}{2\pi}\int B.
\ena
The effective D-brane action for noncommutative SYM is given by
\bea
L = - \frac{1}{G_s (2\pi)^3 \a'^2}\sqrt{-\det(G+2\pi \a' {\hat F})},
\ena
where $G_s$ is the open string coupling defined by~\cite{SW}
\bea
G_s = g_s \left(\frac{\det (g+2\pi\a' B)}{\det g}\right)^{1/2}.
\ena

Expanding the action, we find
\bea
L = - \frac{\sqrt{-g_{00}g_{33}}}{(2\pi)^2g_{s}'{\a'}^2}|B|- \frac{1}{2}
 \frac{\sqrt{-g_{00}g_{33}}}{(2\pi)^4g_{s}'} \frac{k_{1}k_{2}}{|B|}
 - \frac{\sqrt{-\det G}}{4g_{YM}^2}\hat{F}_{G}^{2} + O(\a'),
\ena
where $g_{YM}^2 = 2\pi g_s, g_s = \a' g_s'$ and we have assumed a diagonal
metric $g = {\rm diag.}(g_{00},g_{11},g_{22},g_{33})$ and have written
$g_{11}=\a'^2 k_1, g_{22}=\a'^2 k_2$.
The first term is of order ${\a'}^{-2}$ and gives an infinite contribution
in the $\a'\to 0$ limit. This term is equal to $\mu_{D1}N_{D1}
\sqrt{-g_{00}g_{33}}$. This is the contribution of D1-branes with worldvolume
in the (03) directions. Thus the introduction of $B$ field indeed gives
the correct D1-brane tension. Since a nonvanishing field strength is
equivalent to introducing $B$ field, this suggests that the above solution
with constant field strength can be interpreted as a D1-brane.

The second term is finite in the $\a'\to 0$ limit.
This term gives a contribution of the background charge to the low-energy
($\a'\to 0$) field theory. It can be written as $\frac{1}{4}\frac{\sqrt{
-\det G}}{g_{YM}^2}(\t^{-1})_{G}^2$. The third term is the NCSYM Lagrangian.
In these expressions, the suffix $G$ indicates that the contraction of the
Lorentz indices are made by the open string metric~\p{osm}. These terms can
be combined into
\bea
- \frac{\sqrt{-{\rm det}G}}{4 g_{YM}^2} (\hat{F} -\t^{-1})^2_G,
\ena
which gives the effective action of the NCSYM on the brane.

\section*{Acknowledgements}

One of us (N.O.) would like to thank Paul K. Townsend for valuable
discussions at SI2000. Preliminary results were reported at IIAS (Kyoto)
in June and also at APCTP-Yonsei summer workshop (Seoul) in August.
He thanks B.-H. Lee, K. Lee and S. Terashima for useful discussions on the
latter occasion, and the organizers for hospitality.

This work was supported in part by Grants-in-Aid for Scientific Research
Nos. 99020, 12640270 and in part by Grant-in-Aid on the Priority Area:
Supersymmetry and Unified Theory of Elementary Particles.

\newcommand{\NP}[1]{Nucl.\ Phys.\ {\bf #1}}
\newcommand{\PL}[1]{Phys.\ Lett.\ {\bf #1}}
\newcommand{\CQG}[1]{Class.\ Quant.\ Grav.\ {\bf #1}}
\newcommand{\CMP}[1]{Comm.\ Math.\ Phys.\ {\bf #1}}
\newcommand{\IJMP}[1]{Int.\ Jour.\ Mod.\ Phys.\ {\bf #1}}
\newcommand{\JHEP}[1]{J.\ High\ Energy\ Phys.\ {\bf #1}}
\newcommand{\PR}[1]{Phys.\ Rev.\ {\bf #1}}
\newcommand{\PRL}[1]{Phys.\ Rev.\ Lett.\ {\bf #1}}
\newcommand{\PRE}[1]{Phys.\ Rep.\ {\bf #1}}
\newcommand{\PTP}[1]{Prog.\ Theor.\ Phys.\ {\bf #1}}
\newcommand{\PTPS}[1]{Prog.\ Theor.\ Phys.\ Suppl.\ {\bf #1}}
\newcommand{\MPL}[1]{Mod.\ Phys.\ Lett.\ {\bf #1}}
\newcommand{\RMP}[1]{Rev.\ Mod.\ Phys.\ {\bf #1}}

\end{document}